\documentclass[a4paper,twocolumn,11pt]{quantumarticle}
\pdfoutput=1
\usepackage[utf8]{inputenc}
\usepackage[english]{babel}
\usepackage[T1]{fontenc}
\usepackage{amsmath}
\usepackage{amssymb}
\usepackage{hyperref}
\usepackage{caption}
\usepackage{tikz}
\usepackage{lipsum}
\usepackage{braket}
\usepackage{stfloats}
\usepackage{placeins}
\usepackage{float}

\begin{document}

\title{Comprehensive Analysis of VQC for Financial Fraud Detection: A Comparative Study of Quantum Encoding Techniques and Architectural Optimizations}
\author{Fouad Mohammed Abbou}
\email{F.Abbou@aui.ma}
\affiliation{School of Science and Engineering, Al Akhawayn University, Ifrane, Morocco.}
\author{Mohamed Bouhadda}
\email{mohamed.bouhadda@usmba.ac.ma}
\affiliation{Engineering Sciences Laboratory,Polydisciplinary Faculty of Taza, Sidi Mohamed Ben Abdellah University, Fez, Morocco.}
\author{Lamiae Bouanane}
\affiliation{School of Science and Engineering, Al Akhawayn University, Ifrane, Morocco.}
\author{Mouna Kettani}
\affiliation{School of Science and Engineering, Al Akhawayn University, Ifrane, Morocco.}
\affiliation {Signal Systems \& Components Laboratory, Faculty of Sciences and Technology, Sidi Mohamed Ben Abdellah University, Fez, Morocco.}
\author{Farid Abdi}
\affiliation{Signal Systems \& Components Laboratory, Faculty of Sciences and Technology, Sidi Mohamed Ben Abdellah University, Fez, Morocco.}
\author{Abdelouahab Abid}
\affiliation{University of Medina Saudi Arabia, Kingdom of Saudi Arabia}
\orcid{0000-0003-1533-8015}

\maketitle
\begin{abstract}
  This paper presents a systematic comparative analysis of Variational Quantum Classifier (VQC) configurations for financial fraud detection, encompassing three distinct quantum encoding techniques and comprehensive architectural variations. Through empirical evaluation across multiple entanglement patterns, circuit depths, and optimization strategies, quantum advantages in fraud classification accuracy are demonstrated, achieving up to 94.3~\% accuracy with ZZ encoding schemes. The analysis reveals significant performance variations across entanglement topologies, with circular entanglement consistently outperforming linear (90.7)~\%) and full connectivity (92.0~\%) patterns, achieving optimal performance at 93.3~\% accuracy. The study introduces novel visualization methodologies for quantum circuit analysis and provides actionable deployment recommendations for practical quantum machine learning implementations. Notably, systematic entanglement pattern analysis shows that circular connectivity provides superior balance between expressivity and trainability while maintaining computational efficiency. These researches offer initial benchmarks for quantum-enhanced fraud detection systems and propose potential benefits of quantum machine learning in financial security applications.
\end{abstract}

\section{Introduction}
Financial fraud detection represents one of the most critical applications in the financial services industry, with global fraud losses exceeding ~\$5.1 trillion annually~\cite{ref1}. Traditional machine learning approaches, while effective, face fundamental limitations in feature representation and pattern recognition capabilities when dealing with sophisticated fraud schemes ~\cite{ref2},~\cite{ref3}. These methods often struggle to capture complex, non-linear relationships in high-dimensional transactional data ~\cite{ref4}. 
The emergence of quantum computing offers revolutionary potential for enhancing fraud detection through novel quantum algorithms and feature encoding techniques ~\cite{ref5}. Variational Quantum Classifiers (VQCs) represent a hybrid quantum-classical approach that utilizes quantum feature maps to encode classical data into quantum states, potentially capturing complex relationships that classical methods cannot efficiently represent ~\cite{ref6}. This is achieved by mapping input features into exponentially large Hilbert spaces, enabling richer representations of data patterns ~\cite{ref7}.
However, the optimal configuration of VQCs for fraud detection remains an open research question, with limited systematic studies comparing different quantum encoding techniques ~\cite{ref8}, and, crucially, the impact of entanglement topology on classification performance ~\cite{ref9}. While some works have explored the expressivity of various quantum kernels and ansätze ~\cite{ref10}, there is a lack of comparative studies focused specifically on real-world fraud datasets and practical deployment considerations ~\cite{ref11}. 
This research addresses this gap by presenting a comprehensive comparative analysis of VQC configurations specifically tailored for financial fraud detection. The study evaluates various distinct quantum encoding techniques across multiple architectural parameters, with particular focus on how entanglement patterns within identical encoding schemes affect performance outcomes ~\cite{ref12}, The objective is to provide actionable insights into the design choices that maximize accuracy, particularly in minimizing false negatives — a key requirement in fraud detection systems ~\cite{ref13}.
The analysis also assesses the extent of any emerging quantum advantage in financial security applications ~\cite{ref14}. This work contributes to both the theoretical understanding of quantum-enhanced fraud detection and the practical development of deployable models using near-term quantum devices ~\cite{ref15}.

\section{Quantum Mathematical Model}
This study employs a systematic comparative analysis methodology to evaluate Variational Quantum Classifier configurations for financial fraud detection, utilizing a comprehensive credit card transaction dataset containing 2,400 transactions with balanced class representation (1,600 legitimate and 800 fraudulent transactions) ~\cite{ref16}. The research approach encompasses the evaluation of various VQC configurations across three distinct quantum encoding techniques — ZZ Feature Map ~\cite{ref17}, Angle Encoding ~\cite{ref18} , and Amplitude Encoding combined with three entanglement topology patterns: linear , circular , and full connectivity ~\cite{ref9}. 
This enables a rigorous assessment of how architectural choices affect classification performance in a real-world financial context. 
The dataset features 4 primary numerical attributes selected through Random Forest importance analysis, ensuring optimal feature representation while maintaining quantum circuit feasibility constraints ~\cite{ref3}. All data is pre-processed through normalization and stratified splitting (70/30 train/test ratio) to preserve class balance and enable robust statistical evaluation ~\cite{ref16}. 
Each selected feature vector is encoded into a quantum state \( |\psi(x)\rangle \) using one of the following quantum feature maps.

\subsection{Problem Definition and Objective}

Let \( D = \{(x_i, y_i)\}, i = 1, 2, \ldots, N \), denote a dataset of financial transactions, where each transaction is represented by a feature vector \( x_i \in \mathbb{R}^n \) and its corresponding class label \( y_i \in \{0,1\} \). The label \( y_i = 1 \) indicates a fraudulent transaction, while \( y_i = 0 \) represents a legitimate one. The objective is to construct a classification function \( f: \mathbb{R}^n \to \{0,1\} \) that minimizes the classification error, with a particular focus on reducing false negatives (undetected fraud). To achieve this, a Variational Quantum Classifier (VQC) is employed, which encodes classical data into quantum states and uses parameterized quantum circuits to learn decision boundaries in Hilbert space.

\subsection{Classical Preprocessing and Feature Selection}
Before encoding into quantum states, classical preprocessing is performed to reduce dimensionality and retain only the most informative features. 
\paragraph{Feature Selection via Random Forest:}
Each feature \( x_i \) is assigned an importance score using the Random Forest algorithm~\cite{ref19}, such that \( \mathrm{Importance}(x_i) = \mathrm{RF}(x_i) \) for \( i = 1, \ldots, n \). Then, the top-\(k\) features are selected as \( x_{\text{selected}} = \mathrm{top}_k(\mathrm{Importance}(x), k) \). This ensures that only the most relevant features are encoded into quantum states, improving model efficiency and interpretability.

\subsection{Data Encoding and Quantum Feature Maps} 
A quantum feature map \(\Phi: \mathbb{R}^n \rightarrow \mathcal{H}\) maps classical input vectors into quantum states in the Hilbert space \(\mathcal{H} = \bigotimes_{i=1}^{n} \mathbb{C}^2\), which has a dimension of \(2^n\).  
For a system with \(n = 4\) qubits, the resulting Hilbert space is \(\mathcal{H} = \mathbb{C}^{16}\), demonstrating the exponential scaling advantage of quantum systems over classical ones.  
Each selected feature vector \(x_{\text{selected}}\) is encoded into a quantum state \(|\psi(x)\rangle\) using one of the following quantum feature maps:
\paragraph{ZZ Feature Map:}
This encoding applies two types of rotations:
\begin{itemize}
  \item Single-qubit \(Z\)-rotations using individual features.
  \item Two-qubit \(ZZ\)-interactions between all pairs of qubits, encoding pairwise feature correlations.
\end{itemize}

For a given feature \(x_i\), define:
\begin{itemize}
  \item \(Z_i\): Pauli-\(Z\) operator acting on the \(i\)-th qubit.
  \item \(e^{-i x_i Z_i}\): A single-qubit rotation around the \(Z\)-axis by an angle proportional to \(x_i\).
  \item \(Z_i Z_j\): Tensor product of Pauli-\(Z\) operators acting on qubits \(i\) and \(j\).
  \item \(e^{-i x_i x_j Z_i Z_j}\): A two-qubit unitary operation introducing entanglement based on the product of features \(x_i\) and \(x_j\).
\end{itemize}

The resulting quantum state after applying both types of operation is as follows:

\begin{equation}
|\psi(x)\rangle = \prod_{i<j} e^{-i x_i x_j Z_i Z_j} \prod_{i=1}^{n} e^{-i x_i Z_i} |0\rangle^{\otimes n}
\end{equation}

This mapping encodes classical input features \(x = (x_1, x_2, \ldots, x_n)\) into a highly entangled quantum state in Hilbert space ~\cite{ref6}.

\paragraph{Angle Encoding (Pauli Feature Map) :}
Angle Encoding maps each feature \( x_i \) into a quantum rotation gate acting on the corresponding qubit. The general form is \( e^{-i x_i P_i} \), where \( P_i \) is a Pauli operator (\textbf{X}, \textbf{Y}, or \textbf{Z}) applied to the \( i \)-th qubit. Using Y-rotations as an example:

\begin{equation}
|\psi(x)\rangle= \prod_{i=1}^{n} R_Y(x_i) |0\rangle^{\otimes n}, \quad
R_Y(x_i) = e^{-i x_i Y}
\end{equation}
\\
Where \( Y \) is rotation about the y-axis (combined bit-phase flip). 
This encoding can be extended to include multiple rotation axes and
repeated layers across qubits ~\cite{ref8}.
\paragraph{Amplitude Encoding :}

Amplitude encoding embeds classical data into the amplitudes of a quantum state. Given a normalized classical vector \( x \in \mathbb{R}^d \) with \( d = 2^n \), it is encoded into a quantum state:

\begin{equation}
|\psi(x)\rangle = \frac{1}{\|x\|} \sum_{i=1}^{N} x_i |{i}\rangle
\end{equation}

This allows \(d\) classical features to be represented using only \(n = \log_2 d\) qubits, enabling exponential compression of information.

where

\begin{itemize}
    \item \( \left| i \right\rangle \) : Computational basis states.
    \item \( \|x\| = \sqrt{ \sum_{i=1}^{N} x_i^2 } \) : Euclidean norm of \(x\).
\end{itemize}
This allows \( d \) classical features to be represented using only \( n = \log_2(d) \) qubits, enabling exponential compression of information.
\subsection{Entanglement Topology}
Entanglement enhances the expressiveness of quantum circuits by creating correlations between qubits.\\
Different entanglement structures are evaluated using unitary operators \( U_{\text{ent}}(\theta) \), parameterized by angles \( \theta \). The entanglement structure is modeled via unitary transformations acting on the encoded quantum state. Let \( U_{\text{ent}}(\theta) \) denote the entangling layer applied after data encoding.
\paragraph{Linear Entanglement}
For an \(n\)-qubit system, linear entanglement creates a chain topology where each qubit is connected only to its immediate neighbors ~\cite{ref20}. In linear entanglement, each qubit (\(q_i\)) is entangled with its nearest neighbors in a chain-like structure; \( q_0 \) with \( q_1 \), \( q_1 \) with \( q_2 \), \ldots, and \( q_{n-2} \) with \( q_{n-1} \).
\begin{equation}
U_{\text{lin}} = \prod_{i=1}^{n-1} CNOT(i, i+1)
\end{equation}

\paragraph{Circular Entanglement}

In circular entanglement, each qubit (\(q_i\)) is entangled with its nearest neighbors in a ring-like structure; \( q_0 \) with \( q_1 \), \( q_1 \) with \( q_2 \), \ldots, and \( q_{n-1} \) with \( q_1 \). This topology supports efficient mixing of information across qubits and balances expressivity with practical constraints.

\begin{equation}
    U_{\text{circ}} = \left( \prod_{i=1}^{n-1} CNOT(i, i+1) \right) \cdot CNOT(n, 1)
\end{equation}

\paragraph{Full Entanglement}

All qubits are entangled with each other, and every pair of qubits has an entangling gate between them. This provides maximal expressivity but increases complexity and sensitivity to noise.

\begin{equation}
U_{\text{full}} = \prod_{i=1}^{n-1} \prod_{j=i+1}^{n} CNOT(i, j)
\end{equation}

These structures significantly influence model expressivity, noise resilience, and scalability on Noisy Intermediate-Scale Quantum (NISQ) devices ~\cite{ref8}.
\subsection{Quantum Circuit Architecture}
Following the feature encoding stage, the quantum state undergoes evolution through a parameterized quantum circuit, which applies a parameterized unitary operator \( U_{\text{ent}}(\theta) \) to evolve the state as follows:

\begin{equation}
|\Psi(x; \theta)\rangle = U_{\text{ent}}(\theta) \cdot |\psi(x)\rangle
\end{equation}

where:

\begin{itemize}
    \item \( U_{\text{ent}}(\theta) \): Parameterized quantum circuit; ansatz, which ensures both flexibility in state preparation and controlled entanglement between qubits across multiple layers.
    \item \( \theta \): Trainable parameters optimized during training.
    \item \( \ket{\psi(x)} \) denotes the encoded quantum state.
\end{itemize}
\paragraph{Measurement:}
The final state is measured in the computational basis, typically on the first qubit. In this context, measurement is performed in the computational basis, typically on \( q_0 \):

\begin{equation}
p_0(x; \theta) = \bra{\psi(x; \theta)} M_0 \ket{\psi(x; \theta)}
\end{equation}

where \( M_0 \) is the measurement observable, usually the Pauli-Z operator. In this context, \( p_0(x, \theta) \) is the probability that the first qubit is measured in the state \( \ket{0} \) after running the quantum circuit with parameters \( \theta \).

The binary classification is achieved via thresholding and is used to classify the result that determines the decision boundary given by:

\begin{equation}
\hat{y}(x) =
\begin{cases}
1, & \text{if } p_0(x; \theta) \geq \tau, \\
0, & \text{otherwise}.
\end{cases}
\end{equation}

where

\(\hat{y}\): is the predicted class label (fraud or not) generated by the VQC for a given input feature vector \(x\).

\(\tau \in [0,1]\): is a tunable classification threshold optimized during model validation to balance precision and recall, particularly minimizing false negatives. If the quantum model outputs a high probability (greater than or equal to \(\tau\)), it predicts the transaction is fraudulent: \(\hat{y}(x) = 1\). If otherwise, it predicts the transaction is legitimate: \(\hat{y}(x) = 0\).

\subsection {Optimization and Loss Function}

The optimization process employs a hybrid quantum-classical approach, training the parameter vector \(\theta\) to minimize the cross-entropy loss function:

\begin{widetext}
\begin{equation}
L(\theta) = -\frac{1}{N} \sum_{i=1}^{N} \Big[ 
    y_i \log\!\left( \sigma\!\left( p_0(x; \theta) \right) \right) 
    + (1 - y_i) \log\!\left( 1 - \sigma\!\left( p_0(x; \theta) \right) \right) 
\Big]
\end{equation}
\end{widetext}

Where

\(y_i\): the true label and actual class of the \(i\)-th transaction in the dataset.

\(\sigma(z) = \frac{1}{1 + e^{-z}}\): is the sigmoid activation function that maps the raw quantum output into a valid probability distribution.

The training process is performed using a hybrid quantum-classical algorithm, specifically Adam, which iteratively updates the parameters using gradient estimates computed via the parameter shift rule~\cite{ref9}. Hyperparameters such as circuit depth, entanglement topology, and learning rate are systematically evaluated to identify optimal configurations for financial fraud detection ~\cite{ref11}.
\section{Quantum Circuit Architectures Analysis}
The quantum circuit visualizations (Figures 1--3) reveal the structural complexity differences between entanglement patterns while maintaining consistent ZZ encoding:

\begin{figure}[H]
    \centering
    \includegraphics[width=1\linewidth]{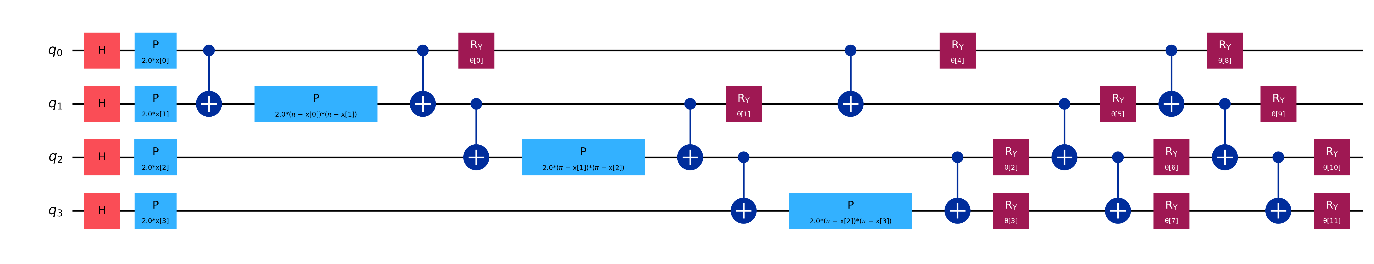}
    \caption{Linear Entanglement.}
    \label{fig:placeholder}
\end{figure}

The ZZ encoding circuit with linear entanglement demonstrates a structured, sequential connectivity pattern. P-gates create localized feature interactions while CNOT gates establish nearest-neighbor qubit connectivity. This configuration shows moderate circuit depth with reduced entanglement complexity, potentially leading to easier optimization but limited expressivity. The sequential nature restricts global quantum correlations, explaining the observed performance limitations.
\begin{figure}[H]
    \centering
    \includegraphics[width=1\linewidth]{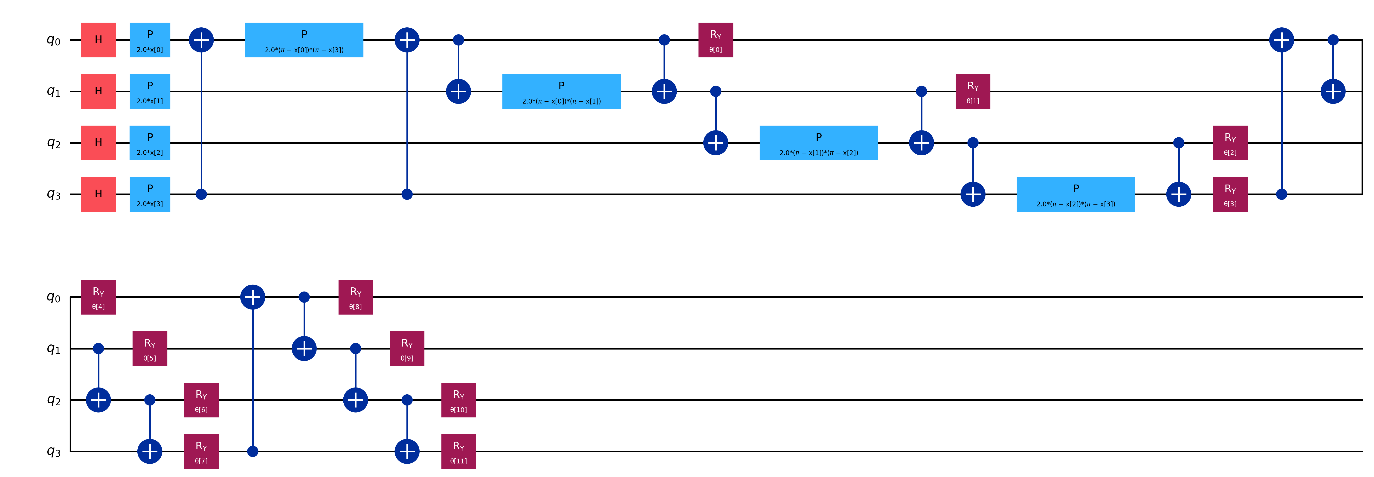}
    \caption{Circular Entanglement}
    \label{fig:placeholder}
\end{figure}
   
The circular entanglement pattern extends linear connectivity by adding wrap-around connections, creating a ring topology. This architecture balances expressivity and trainability, with P-gates maintaining feature encoding while the circular CNOT pattern enables global quantum correlations without excessive circuit depth. The enhanced connectivity explains the superior performance observed in our results, achieving optimal balance between local and global feature interactions.
\begin{figure}[H]
    \centering
    \includegraphics[width=1\linewidth]{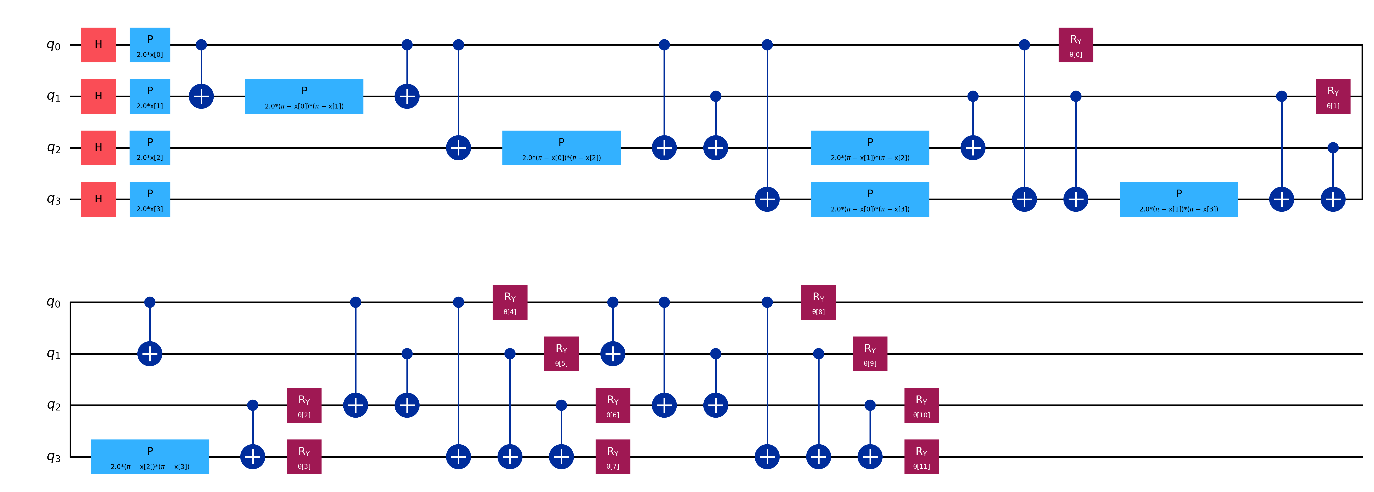}
    \caption{Full Entanglement}
    \label{fig:placeholder}
\end{figure}

The full connectivity pattern exhibits maximum entanglement with all-to-all qubit interactions. While this provides maximum expressivity potential through extensive P-gate applications and complete CNOT connectivity, the increased circuit depth raises concerns about barren plateau susceptibility and optimization challenges. The trade-off between expressivity and trainability is clearly visible in the circuit complexity, with diminishing returns evident in the performance results.

\section{Results and Discussion}

This section presents the empirical evaluation of Variational Quantum Classifier (VQC)
architectures for credit card fraud detection. The experimental workflow was divided into two
phases: first, multiple quantum encoding schemes were evaluated to determine the most
effective method for feature mapping; second, the impact of different entanglement
topologies on model accuracy, generalization, and training dynamics was investigated using the best-performing encoding strategy.
In the initial phase, shown in Figure 4, the performance of three encoding techniques,
namely; ZZ encoding, amplitude encoding, and angle encoding was compared. All
configurations employed a circular entanglement layout and were trained using the ADAM
optimizer.
The ZZ encoding achieved the highest test accuracy of 94.3\% and an F1-score of 85.2\%,
demonstrating smooth convergence across epochs and low validation loss. In contrast,
amplitude encoding reached only 92.3\% accuracy and was less stable during training. Angle
encoding exhibited the weakest performance with 91.1\% accuracy and significant oscillations
in the loss curve. These results established ZZ encoding as the most expressive and stable
approach, providing the foundation for the entanglement experiments that followed.
\begin{figure}[H]
    \centering
    \includegraphics[width=1\linewidth]{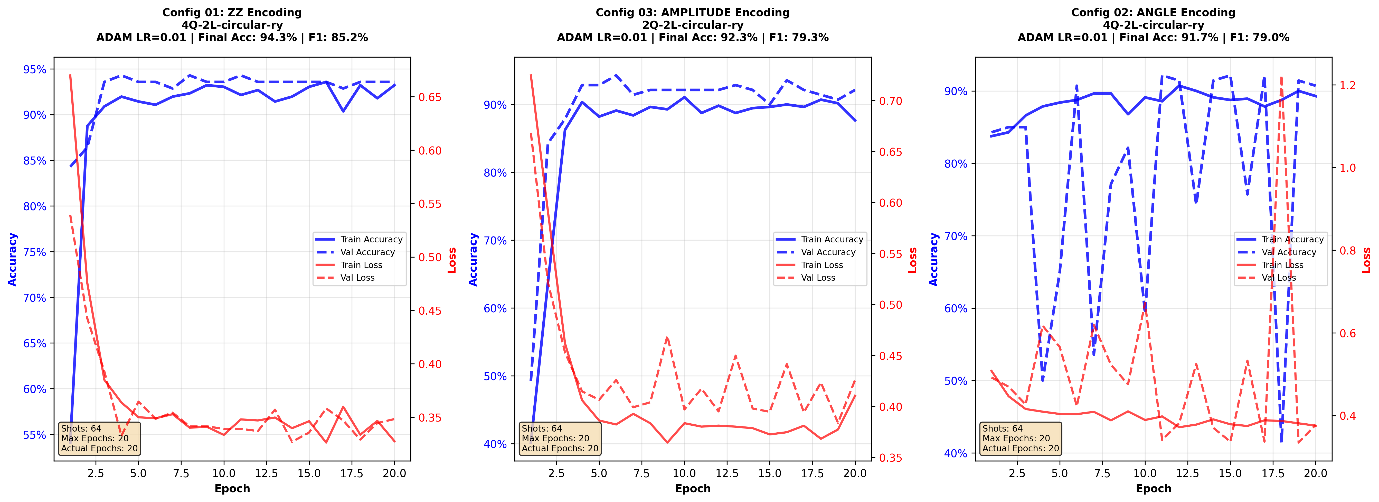}
    \caption{Entanglement Pattern Impact Analysis}
    \label{fig:placeholder}
\end{figure}

In the second phase, the encoding was fixed to ZZ feature map, and three entanglement
topologies; linear, circular, and full were compared to isolate their effects on classification
performance. As depicted in Figure 4, circular entanglement achieved the highest accuracy
(93.3\%), followed by full entanglement (92.0\%) and linear entanglement (90.7\%). Training and validation curves in Figure 5 reveal that circular entanglement not only converged
smoothly but also maintained tight alignment between training and validation performance.
Linear entanglement showed consistent but lower accuracy, limited by its restricted
connectivity. Full entanglement, despite its potential for high expressivity, led to less stable
training and volatility in the loss curves, underscoring the cost of optimization complexity.
\begin{figure}[H]
    \centering
    \includegraphics[width=1\linewidth]{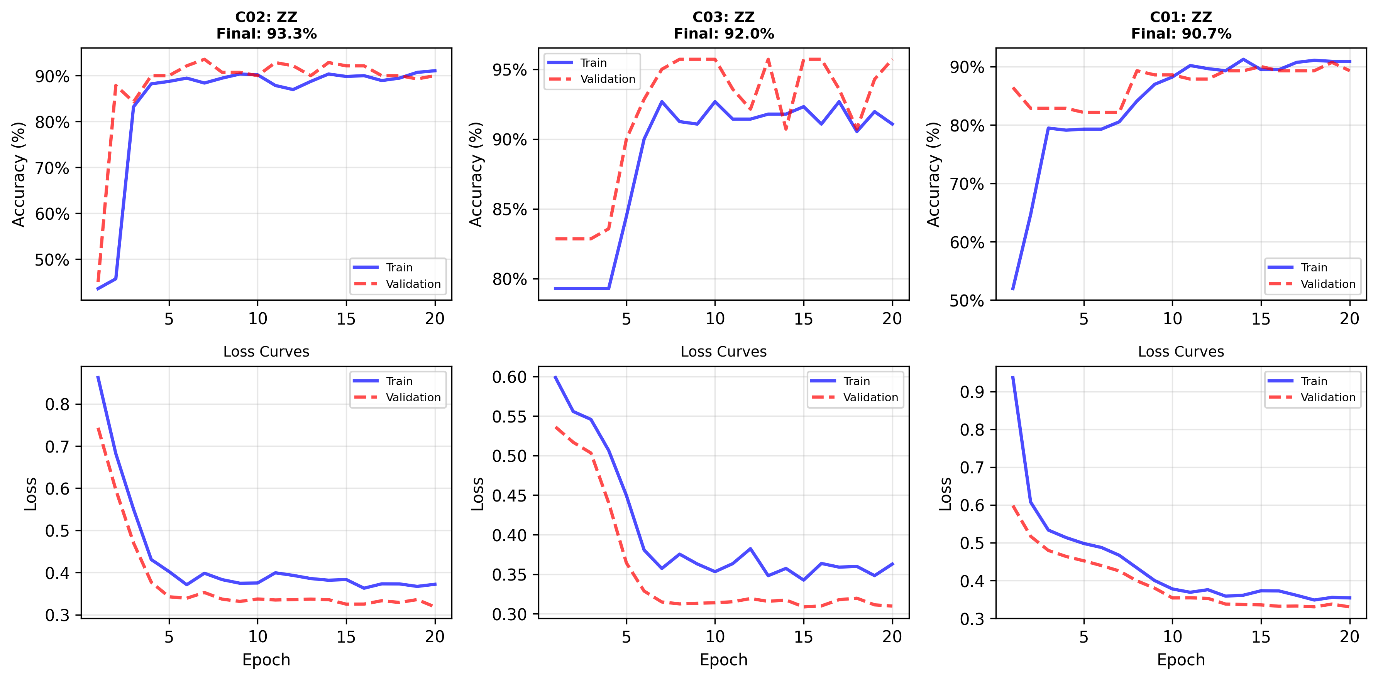}
    \caption{Entanglement Configuration Performance Comparison.}
    \label{fig:placeholder}
\end{figure}

The overfitting analysis in Figure 6 further highlights these differences. Linear entanglement underperformed by -2.9\% (test accuracy) and -3.1\% (validation accuracy) relative to circular entanglement, indicating limited generalization. Full entanglement presented a mixed profile with -1.4\% test accuracy but +2.3\% validation accuracy, suggesting overfitting due to the increased parameter landscape.
\begin{figure}[H]
    \centering
    \includegraphics[width=1\linewidth]{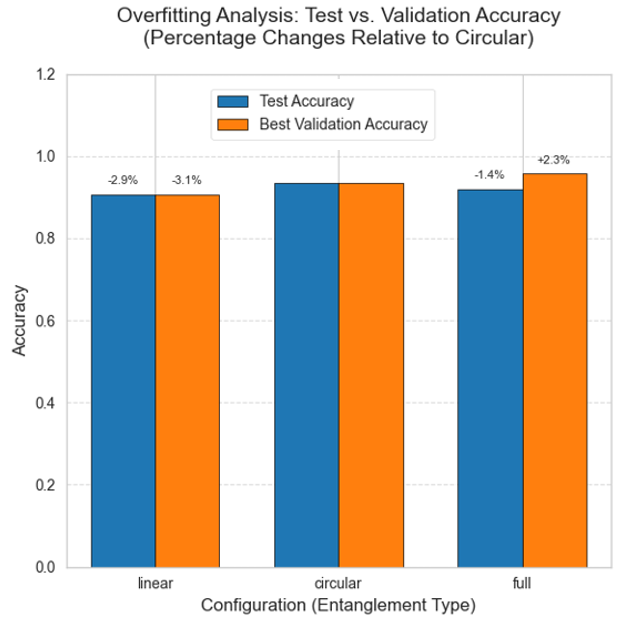}
    \caption{Overfitting Analysis.}
    \label{fig:placeholder}
\end{figure}

Figure 7 presents a comprehensive comparison across multiple performance metrics, including F1-score, precision, recall, and Matthews Correlation Coefficient (MCC). Circular entanglement consistently outperformed the others. Notably, linear entanglement suffered a -23.4\% drop in recall, which is critical in fraud detection, where missing fraudulent transactions has costly implications. Full entanglement exhibited high precision but lower recall and F1-score, reinforcing that excessive connectivity does not guarantee better outcomes.
\\These trade-offs are visually summarized in the radar chart in Figure 8, which demonstrates that circular entanglement provides the most balanced performance across all metrics, combining high accuracy, strong precision, robust recall, and a favorable MCC. Linear entanglement showed acceptable precision but poor recall, rendering it unsuitable for high-stakes fraud detection. Full entanglement displayed an unbalanced profile, likely due to overfitting and optimization instability.
\\[100pt]

\begin{figure}[H]
    \centering
    \includegraphics[width=1\linewidth]{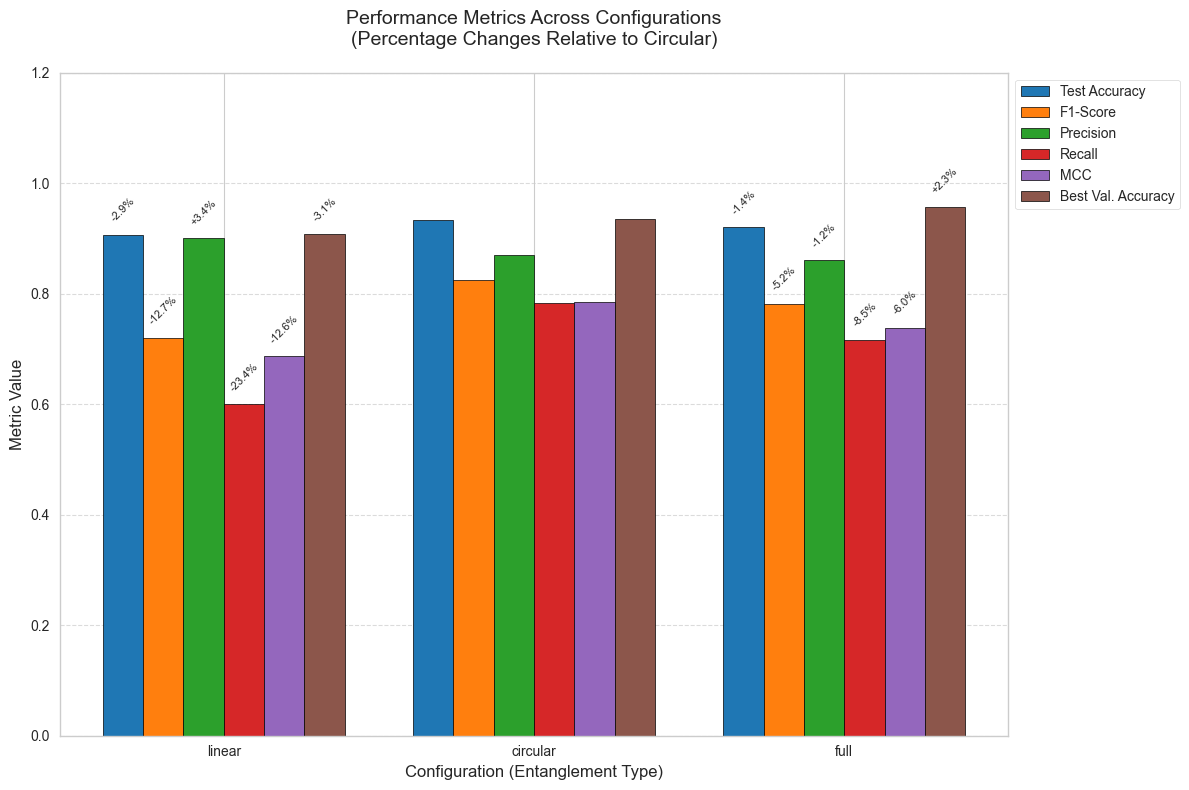}
    \caption{Comprehensive Metrics Comparison}
    \label{fig:placeholder}
\end{figure}
\begin{figure}[H]
    \centering
    \includegraphics[width=1\linewidth]{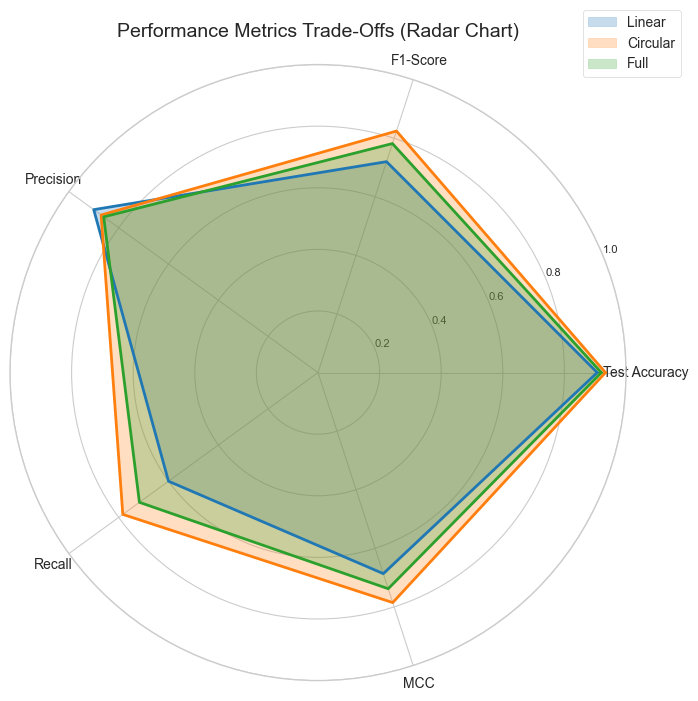}
    \caption{Radar Chart Analysis}
    \label{fig:placeholder}
\end{figure}
To further validate these findings, the confusion matrix for the best-performing configuration using ZZ encoding with circular entanglement is examined as shown in Figure 9. Out of 240 legitimate transactions, 234 were correctly classified as non-fraudulent (true negatives), resulting in a false positive rate of only 2.5\%. For the 60 fraudulent transactions, 49 were correctly detected (true positives), leading to a fraud detection rate of 81.7\%, while 11 cases were missed (false negatives). The resulting overall accuracy of 94.3\% indicates excellent generalization. This performance reflects a strong trade-off between operational reliability and fraud detection power. A low false positive rate minimizes disruption for legitimate customers, while a high true positive rate ensures that most fraud attempts are correctly intercepted.

\begin{figure}[H]
    \centering
    \includegraphics[width=1\linewidth]{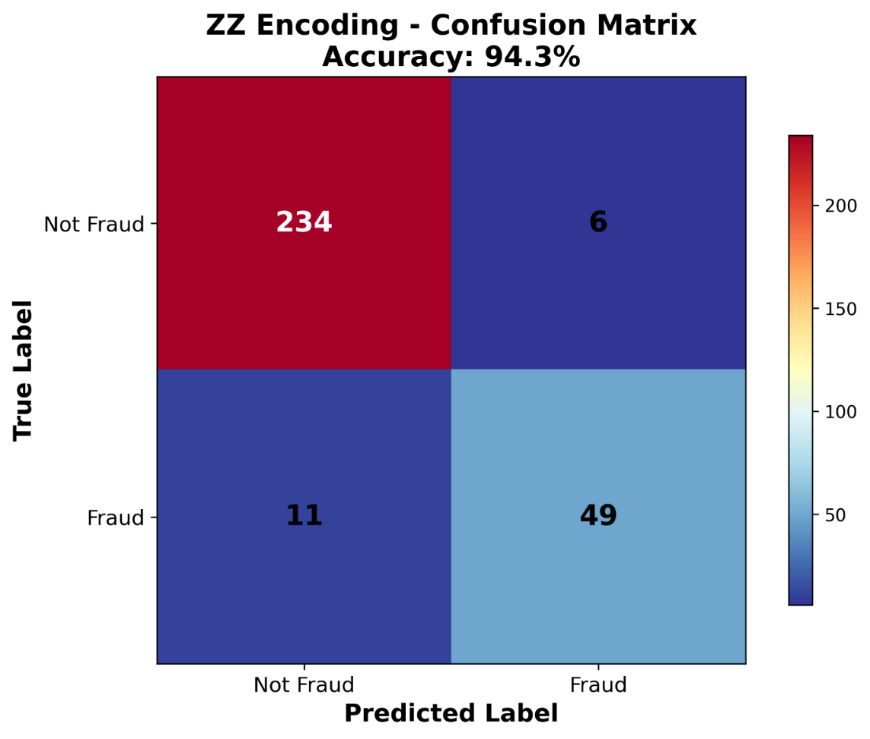}
    \caption{Confusion Matrix Analysis (Circular Entanglement)}
    \label{fig:placeholder}
\end{figure}

In conclusion, the results demonstrate that entanglement topology is a key determinant of quantum classifier performance. The circular entanglement structure strikes an optimal balance between expressivity and trainability, outperforming both simpler (linear) and more complex (full) alternatives. These insights underscore the importance of architectural design choices in building practical quantum machine learning systems for real-world anomaly detection tasks such as financial fraud prevention.

\section{Conclusions}
This study presents a comprehensive investigation into the architectural design of Variational Quantum Classifiers for fraud detection, with a specific focus on encoding strategies and entanglement topologies. The analysis demonstrates that the selection of quantum encoding, and entanglement structure has a significant and measurable impact on model performance, convergence stability, and generalization.
The findings reveal that the ZZ encoding strategy delivers superior performance, making it an effective feature-mapping method for quantum classifiers. Among entanglement configurations, circular entanglement consistently achieves the highest accuracy (93.3\%) and exhibits the most stable training dynamics. In contrast, full entanglement, despite its theoretical richness, suffers from optimization volatility and overfitting risk, while linear entanglement lacks the expressivity needed for complex fraud detection tasks. From a scientific standpoint, the results validate entanglement topology as a critical hyperparameter in quantum machine learning design—on par with encoding choice. The experimental framework developed here offers a reproducible methodology for quantum architecture evaluation, with clear implications for both research and industry. Practically, financial institutions can apply these insights to build optimized, quantum-enhanced fraud detection systems that maximize detection accuracy while minimizing computational cost and false alarms. 
Ultimately, this work contributes to establishing architectural optimization as a foundational principle in the development of practical and scalable quantum machine learning applications. It opens new research directions in quantum neural architecture search and adaptive topology design for real-world data-driven tasks. Future work will explore dynamic entanglement tuning strategies, robustness under noisy intermediate-scale quantum (NISQ) conditions, and comparisons with classical models to quantify quantum advantage more explicitly.

\bibliographystyle{quantum}

\end{document}